\documentclass[journal]{IEEEtran}

\ifCLASSINFOpdf
\else
   \usepackage[dvips]{graphicx}
\fi
\usepackage{url}

\usepackage{graphicx}
\usepackage{amsmath}

\begin{document}

\title{Pitch Preservation In Singing Voice Synthesis}

\author{Shujun Liu, Hai Zhu, Kun Wang, Haujun Wang 
\thanks{S. Liu and H. Wang are with the College of Geophysics, Chengdu University of Technology, Chengdu 610059, Sichuan, China (e-mail:
suldier@outlook.com; wanghuajun@mail.cdut.edu.cn). }
\thanks{H. Zhu and K. Wang are with the Changhong AI Lab (CHAIR), Sichuan Changhong Electronics Holding Group Co., Ltd. Chengdu 610096, Sichuan, China (e-mail: hai1.zhu@changhong.com; kun1.wang@changhong.com). }}

\markboth{}{}
\maketitle

\begin{abstract}
Suffering from limited singing voice corpus, 
existing singing voice synthesis (SVS) methods that build encoder-decoder neural networks to directly generate spectrogram could lead to out-of-tune issues during inference phase.
To attenuate these issues,
this paper presents a novel acoustic model with independent pitch encoder and phoneme encoder,
which disentangles the phoneme and pitch information from music score to fully utilize the corpus.
Specifically, according to equal temperament theory, the pitch encoder is constrained by a pitch metric loss that maps distances between adjacent input pitches into corresponding frequency multiples between the encoder outputs.
For the phoneme encoder, based on the analysis that same phonemes corresponding to varying pitches can produce similar pronunciations,
this encoder is followed by an adversarially trained pitch classifier  to enforce the identical phonemes with different pitches mapping into the same phoneme feature space. By these means, the sparse phonemes and pitches in original input spaces can be transformed into more compact feature spaces respectively, where same elements cluster closely and cooperate mutually to enhance synthesis quality.
Then, the outputs of the two encoders are summed together to pass through the following decoder in the acoustic model.
Experimental results indicate that the proposed approaches can characterize intrinsic structure between pitch inputs to obtain better pitch synthesis accuracy and achieve superior singing synthesis performance against the advanced baseline system.
\end{abstract}


\IEEEpeerreviewmaketitle

\section{Introduction}

\IEEEPARstart{U}{nlike} text to speech (TTS) which is a task of converting text and prosody into speech voice,
the SVS aims to generate singing voice from music score that contains lyric, pitch and rhythm information.
Recently, neural network has been successfully employed in acoustic model and vocoder to improve SVS effects essentially.
For acoustic modeling, the methods based on DNN \cite{nishimura2016singing} and CNN \cite{2020Fast} have achieved higher synthesis quality than traditional HMM \cite{saino2006hmm} based methods.
By introducing attention or self-attention mechanism,
autoregressive models \cite{2019Adversarially, blaauw2020sequence} have gained significant improvement in naturalness aspect.
Compared with the autoregressive models, 
which may cause large time consuming because of temporal dependency when decoding current frame,
the literature \cite{Lu2020} of non-autoregressive counterpart  provides a fast way to generate mel-spectrogram  in parallel.
Meanwhile, vocoders based on neural network, such as WaveNet \cite{oord2016wavenet}, WaveRNN \cite{9362104} and WaveGlow \cite{valle2020mellotron}, are also applied to converting acoustic feature to singing audio,
avoiding artifacts due to signal processing algorithms \cite{oord2016wavenet}.

Nevertheless, it still remains an important and challenging problem existing in the SVS system that the acoustic model cannot produce the pitch of singing voice  precisely,
especially when only rare note pitches in training set are available \cite{Lu2020}. 
It is also referred to as out-of-tune issues.
In general,
existing methods to address this problem can be divided into three categories: data augmentation, pitch normalization and pitch post-processing.
For example, the data augmentation based method of \cite{blaauw2017neural} applies pitch shifting in semitone to both the note pitch of music score and F0 of corresponding audio for each training sample,
resulting in that the pitch shifted pseudo data can sufficiently cover pitch range of original data set,
but inevitably consumes training time in multiples.
The pitch normalization based method of \cite{saino2010singing} considers using the deviation between  LF0 (log F0) and note pitch as an acoustic feature, which is summed with note pitch to feed WORLD vocoder \cite{morise2016world} in inferencing stage.
As a result, 
it produces lower naturalness than neural network based vocoders that take the entire spectrogram as acoustic feature directly,
because the conventional vocoder based on source filter model that decomposes singing signal into three acoustic features (e.g., MGC, BAP and LF0) cannot recover the singing signal without distortion.
Moreover, the pitch post-processing based method of \cite{Yi2019} that modifies the predicted F0 by smoothing according to note pitch may lose vibrato information.
All above mentioned methods do not consider the linear relationship between note pitches in input sequence.
That means, for spectrogram modeling based on encoder-decoder architecture, the acoustic model cannot characterize true pitch distance in spectrogram,
and hence may cause lower pitch accuracy.

To alleviate the out-of-tune issues for SVS system based on spectrogram modeling,
we introduce two encoders to separately represent pitch and phoneme features.
Specifically,
the phoneme encoder with a pitch metric loss,
which constrains geometric distances between the phoneme encoder outputs,
enforces its outputs to form clear clusters according to pitch inputs categories.
On the other hand,
we apply adversarial domain adaption for the phoneme encoder to learn a pitch-independent phoneme representation.
As domain adaption has attracted many research for generating voice in recent years,
various adversarially trained domain classifiers are also designed for different tasks,
such as speaker classifier and tone classifier for multi-speaker cross-lingual TTS \cite{liu2020tone},  language classifier for cross-lingual TTS \cite{xin2020cross},
noise classifier for voice cloning from noise sample \cite{Cong2020}, speaker-singer classifier for cloning speech to singing \cite{9383585},
and singer classifier for multi-singer SVS \cite{Wu2020}.
It is worth mentioning that, to obtain more accurate pitch translation on singing voice conversion task, 
\cite{9054199} alternately updates two adversarial pitch regression losses,
which are negative to each other.
The training strategy forces encoder to generate pitch invariant phoneme representation,
and the removed pitch information in the encoder is compensated by feeding explicit pitch to decoder.
Different from \cite{9054199}, in this paper, we adopt adversarial pitch classifier instead of the pitch regression for convenient training.

\section{Pitch Preservation In Singing Voice Synthesis}

\begin{figure}[t]
  \centering
  \includegraphics[width=\linewidth]{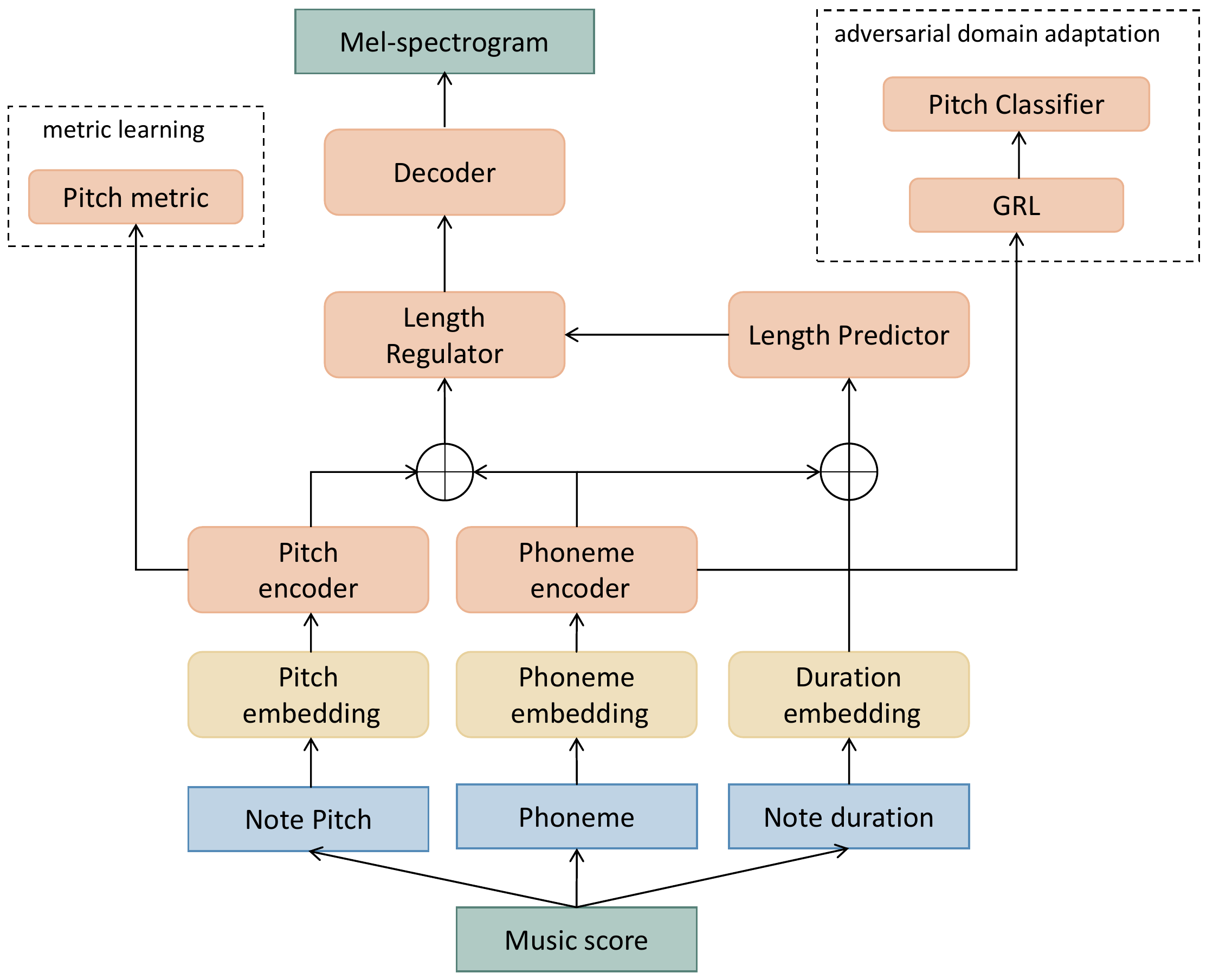}
  \caption{Overview of the proposed architecture with constrained pitch encoder and phoneme encoder.}
  \label{fig:f1}
\end{figure}

In this section, we extend FastSpeech modules \cite{ren2019fastspeech, ren2021fastspeech} to build a non-autoregressive acoustic model for SVS system with two independent encoders as basic architecture drawn in Figure~\ref{fig:f1},
where the proposed methods illustrated in the dotted boxes contain two significant modules:
pitch metric and pitch adversarial classifier.
The pitch metric loss has integrated the geometric distance between pitches,
and the pitch classifier introduces pitch prior into cross entropy weight and GRL (gradient reversal layer) \cite{ganin2015unsupervised} to propagate opposite gradient backward.
Mel-spectrogram is selected as the output of the acoustic model.
In order to adapt the parallel acoustic architecture, we also adopt  non-autoregressive vocoder of Multi-band MelGAN \cite{9383551} for converting mel-spectrogram to waveform in parallel.

\subsection{Separate Encoders For Pitch And Phoneme Representations}
\label{2.1}
As illustrated in Figure~\ref{fig:f1}, the phoneme embedding, note pitch embedding and note duration embedding are extract from music score.
The proposed architecture is different from most of SVS acoustic methods which add or concatenate 
these embeddings together to feed into a common encoder. 
Firstly,
we adopt two encoders with same structure to separately represent phoneme feature and pitch feature based on two major motivations.
One is that the size of pitch vocabulary is comparable to that of phoneme vocabulary because of broad vocal range in singing, so they are also important for representing singing pronunciation. 
And another is that we expect the acoustic model can disentangle pitch representation and phoneme representation to reduce the sparsity of their united representation space.
Then, the sum of the two encoder outputs is passed through the length regulator, which duplicates the phoneme-level feature according to phoneme duration.
Note that in training step the phoneme duration is prepared, but in inference step it derives from the length predictor that takes the sum of phoneme encoder output  and duration embedding as input.
Finally,  
the extended feature is converted to frame-level mel-spectrogram by decoder.
The acoustic model and length predictor are jointly trained  with $\mathcal{L}_1$ loss for mel-spectrogram and $\mathcal{L}_2$ loss for phoneme duration,
\begin{equation}
\mathcal{L}=  \mathcal{L}^{mel}_1 + \mathcal{L}^{dur}_2.
\end{equation} 
\subsection{Pitch Metric}
\label{2.2}
Due to the phenomenon that synthesized singing is inevitably out of tune when limited corpus is used for training,
we consider a solution of introducing the intrinsic relation between input note pitches to acoustic model to alleviate this problem.
Suppose that $p$ denotes pitch ID and $f$ is its corresponding fundamental frequency.
The difference of two pitches, $p_i$ and $p_j$, in semitone and corresponding frequency ratio can be written as follows,
\begin{equation}
  f_{i}=  f_{j} 2 ^ {\frac{1}{12} (p_i - p_j)}.
\end{equation}
It should be noted that there exists addition relationship between $p$ fed to the encoder and corresponding multiplication relationship between $f$ produced by the decoder in our acoustic model.
We attempt to introduce this relationship into our phoneme encoder by 
metric learning \cite{suarez2020tutorial}, which focuses on learning a feature space where distances between features can be distinctly characterized.
Here it is introduced to learn pitch distance on pitch representation space.
To this end, we define the following pitch metric loss for each music score phrase,
\begin{equation}
\label{metric}
\begin{aligned}
  \mathcal{L}_{pm} &= \frac{1}{N}{ \sum_{i=1}^N{ \frac{1}{2k} \sum_{j=-k}^{k}{ |E_{pit}(p_i) - r_{ij}E_{pit}(p_j) |}}},\\
  r_{ij} &=  2 ^ {\frac{1}{12} (p_i - p_j)},
\end{aligned}
\end{equation}
where $E_{pit}$ denotes pitch encoder, $N$ denotes the length of pitch input sequence expanded from score phrase,
$p_i$ denotes the $i$-th pitch ID in the pitch sequence, 
$k$ denotes the count of pitch samples adjacent to the $p_i$ before or after,
and $r$ is the frequency ratio corresponding to the difference between adjacent pitch ID and could be precomputed before training stage.
The pitch metric loss enforces same pitches to cluster together in pitch feature space, which is helpful to sparse pitches.
In addition, the loss  is placed between the pitch encoder and the decoder, desiring that the pitch encoder transforms pitch relationship into frequency relationship so that the decoder can easily learn accurate F0.
The distribution of positive and negative pitch samples depends on whether they are equal to the anchor pitch.
In pop music, the melodies are usually written within 2 octaves and the neighboring notes in one phrase are rarely across one octave,
which means that the pitch encoder tends to embed shorter pitch interval in each score phrase when implementing the proposed strategy of choosing contrast samples.
Hence, compared with choosing contrast samples from other input sequence in a batch,  pitch encoder carrying out the proposed strategy has the advantage of exploiting pitch context information through self-attention of the pitch encoder.

\subsection{Pitch Classifier}
Although the proposed architecture represents phoneme and pitch using two separate encoders to decompose score information,
the phoneme encoder still mixes redundant pitch information derived from the mel-spectrogram by gradient reduction back propagation algorithm in training stage.
To remove the pitch information from phoneme encoder,
we stake gradient reversal layer and pitch classifier after the phoneme encoder sequentially, as demonstrated in Figure~\ref{fig:f1}.
The backward gradient derived from the pitch classifier is set to be its opposite number, 
and the reverse gradient continues to pass back to the phoneme encoder.
Hence, the phoneme encoder can map same phonemes corresponding to different pitch into same feature space,
which is helpful in improving synthesis quality via sharing phoneme representation especially for sparse couples of phoneme and note pitch.

Moreover, note pitch unbalance is another intractable problem.
Because of this, the pitch classifier tends to assign higher probability to classes with larger count of samples,
which may leads to inaccurate results.
To address this problem, we incorporate pitch prior probability into weighted cross entropy \cite{aurelio2019learning} as pitch classifier loss function, defined by
\begin{align}
\mathcal{L}_{pc} &= \frac{1}{N}\sum_{i=1}^N \sum_{j=1}^{M}{-y_j log(C(E_{pho}(t_i)))\lambda_j},\\
\lambda_j &= (\frac{O}{O_j}),
\end{align}
where $N$ denotes the length of pitch input sequence,
$M$ denotes the number of pitch categories,
$y_j$ is  the $j$-th pitch ID label, 
$t_i$ is the $i$-th phoneme ID in phoneme input sequence expanded from the lyric,
$E_{pho}$ is phoneme encoder, 
and $C$ is pitch classifier.
Let $O$ be the total pitch number and $O_j$ be the count of pitch $p_i$ in training set.
The weight  $\lambda_j$ is calculated by the ratio of $O$ to $O_j$ in advance.

\section{Experiments}
\subsection{Dataset}
110 Mandarin pop songs sung by a female singer are recorded in a professional recording environment, covering pitch range from E3 (164.8 Hz) to C6 (1046.5 Hz).
The corresponding music scores are manually annotated in MusicXML format.
Then, the sampling rate of these audios sampled at 48Hz with 16bits is reduced to 24Hz.
The music score and the down-sampled audios are split into score phrases and audio segments of 4515 at the rest locations on music score.
Similar to \cite{Zhang2020}, we prepare paired audio and atonal phoneme sequence  as alignment corpus, which can massively reduces pronunciation units.
After that, the phoneme duration is obtained by implementing the Montreal Forced Aligner \cite{2017Montreal}, a sophisticated audio-text alignment tool developed on HMM algorithm.
Finally, we randomly take ten and five percent of the total songs as testing set and validation set, while the rest as training set.

\subsection{Experimental Configuration}
To investigate the ability of the proposed acoustic model, ablation study that contains the following five models with different architectures and proposed loss function combinations is conducted in this section.
\begin{itemize}
\item \emph{Baseline:}  The Fastspeech for converting  music score to mel-spectrogram.
\item \emph{Proposal:}  The proposed acoustic model with two separate encoders to present pitch and phoneme information respectively.
\item \emph{Pro-pm:} The \emph{Proposal} with pitch metric loss following the pitch encoder to preserve the intrinsic geometry between pitches.
\item \emph{Pro-pc:} The  \emph{Proposal} with pitch classifier following the phoneme encoder to learn pitch-independent phoneme feature.
\item \emph{Pro-pm-pc:} The  \emph{Proposed} system embedding the pitch metric loss and the pitch classifier to improve pitch synthesis accuracy.  
\end{itemize}

In our experiments, only three music properties extracted from music score are used as system inputs, including phoneme, note pitch and note duration.
The phoneme consists of a consonant or a vowel decomposed from Pinyin, which is obtained by converting the Mandarin lyric and manually completing polyphone disambiguation.
The size of phoneme vocabulary is 60, and the range of pitch ID referencing to MIDI standard format is from 52 to 84 owing to limited amount of singing voice corpus.
 In addition, note duration  is represented in second by calculating from time signature and BPM (Beats Per Minute).
Above three information are embedded into 256-dimensional trainable vectors respectively.

The architecture of the \emph{Baseline} follows the setting in \cite{ren2021fastspeech}, but adopt mel-spectrogram  as target acoustic feature.
The remaining parameter configuration is similar to the \emph{Proposal}.
The phoneme encoder and the pitch encoder employ same architecture configuration separately, which  stacks 3 FFT (Feed-Forward Transformer) blocks 
where each block contains multi-head attention with 2 heads and head size of 16 and a 2-layer 1D convolutional network with 3 kernel size.
The  hidden size of each convolutional  layer is 1024 and 512 respectively.
The decoder have the same architecture as the encoder, 
but both of FFT blocks and the number of attention heads in the decoder are  set as 4 and the decoder output size  is 80 for matching mel-spectrogram dimension.
The length predictor is a  2-layer 1D convolutional network with kernel size of 3,
and the input or output size of 256 for each layer.
Besides, dropout probabilities are 0.1, 0.5 and 0.2 for attention layer, length predictor and rest model respectively.

The pitch classifier shown in Figure~\ref{fig:f1}  is composed of a gradient reversal operator and a multi-classifier.
The classifier consists of a 2-layers 1D convolutional network and softmax activation.
ReLU activation and layer normalization follow each network layer.
The output size of the first layer and second layer are all set to be 128,
and the kernel size of 3 is chosen for each layer.
The softmax converts the 128-dimensional vector  into 84-dimensional probability density.

As analyzed in Section \ref{2.2},  choosing $k$ number in Eq. (\ref{metric}) corresponding to neighboring pitches directly  affects pitch representation.
That means, if we select a small value of $k$, the pitch encoder cannot learn global and complete pitch manifolds,
and by contraries, if we set it as a larger value, there will lead to the problems of training consumption and hard convergence.
Fortunately, most out-of-tune singings synthesized deviate from a few semitones,
and the neighboring note pitches in one phrase are rarely across one octave.
In view of above observation, we set $k$ to be 1, selecting two nearest neighbors.

In the training stage,
mini batch size of 32 and Adam optimizer with $\beta_1 = 0.9$, $\beta_2 = 0.98$, $\epsilon = 10^{-6}$ are adopted for optimizing all acoustic models.
Parameters related to their learning rate are selected as follows: initial learning rate of $10^{-3}$, end learning rate of $10^{15}$, decay steps of $10^5$, warmup proportion of $10^{-2}$ and weight decay of $10^{-3}$.
In the training of Multi-band MelGAN, 
hyper-parameters in network and optimizer are configured following \cite{9383551} except that
the learning rate is initially set to be 0, and then reduced by half every 10k iterations.

In order to compare the effects of different systems, a generalized loss is defined as follows,
\begin{equation}
\mathcal{L}_{all} =  \mathcal{L}^{mel}_1 + \mathcal{L}^{len}_2 + \lambda_{1} \mathcal{L}_{pm} + \lambda_{2} \mathcal{L}_{pc}
\end{equation} 
where $ \lambda_{1}$ and  $\lambda_{2}$ are adjustable weights for pitch metric loss and pitch classifier loss respectively.
The last two items are unnecessary for the \emph{Baseline} and the \emph{Proposal}.
The trade off parameters of $(\lambda_1,\lambda_2)$ are manually tuned to the best performance on validation set for the \emph{Pro-pm}, the \emph{Pro-pc} and the \emph{Pro-pm-pc} with $(1, 0)$, $(0, 0.5)$ and (1.1, 0.2) respectively. Specifically, having fixed the $\lambda_1$ with $1$ firstly, we optimize $\lambda_2$ by enumerating array from $0.05$ to $2$.
Alternately, fixing the $\lambda_2$, the optimal $\lambda_1$ is found in the same way.

\subsection{Evaluation}
\subsubsection{Objective Evaluation}

\begin{figure}[t]
  \centering
  \includegraphics[width=\linewidth]{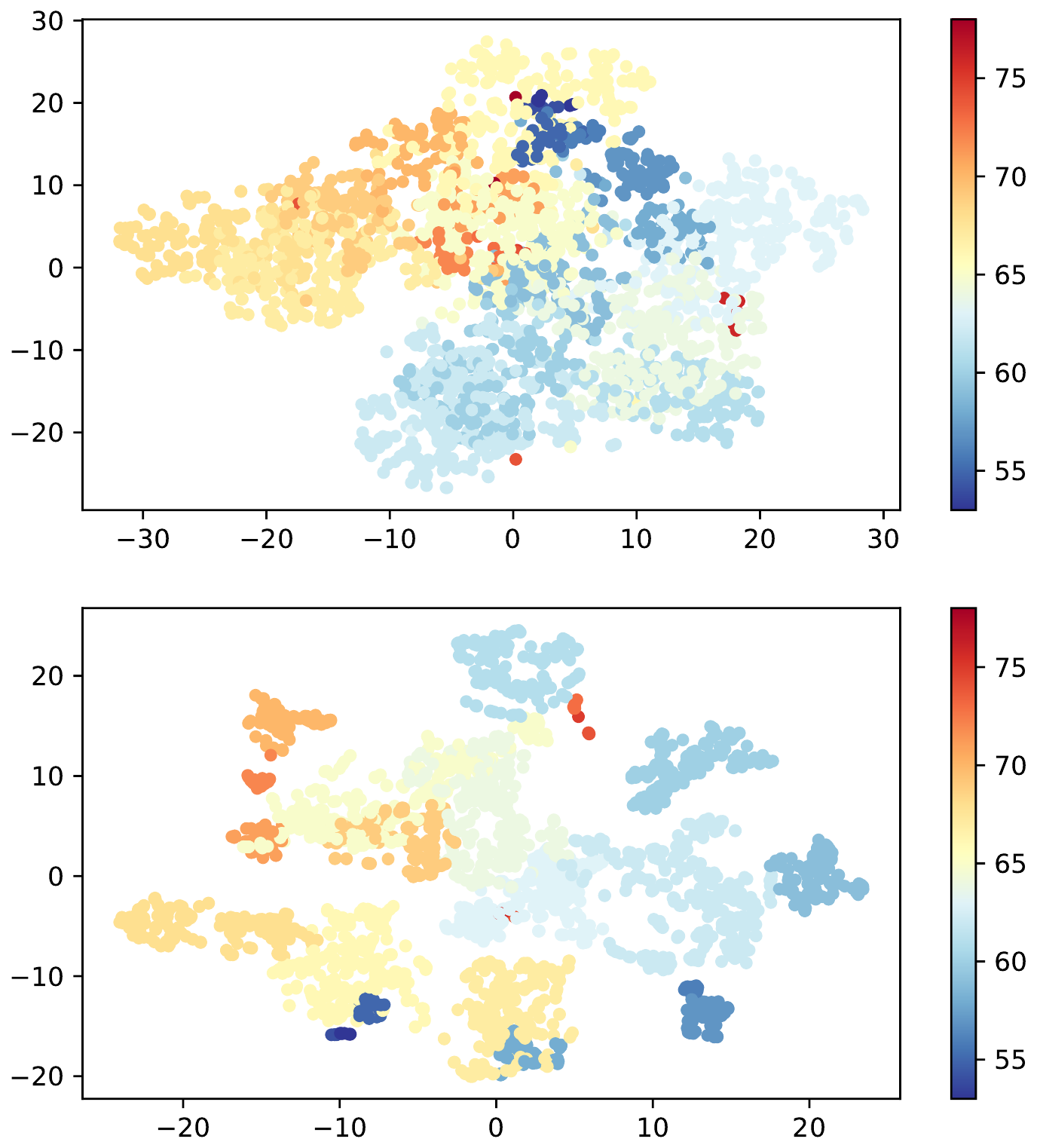}
  \caption{ Visualization of representations outputted from pitch encoders of the \textbf{Proposal} (top) and the \textbf{Pro-pm} (bottom) respectively using t-SNE.}
  \label{fig:f2}
\end{figure}

Three kinds of objective criteria that include F0 root mean square error (RMSE), F0 correlation coefficients and duration accuracy with reference to \cite{9383585} are conducted to evaluate different models.
In order to fairly compare F0 of synthesized singing by each model, we set real duration to all models instead of predicted duration.
As presented in Table~\ref{tab:example}, the \emph{Proposal} achieves better scores of F0 and duration than the \emph{Baseline}.
This may be because the independent encoder representing pitch has stronger ability than the shared encoder
and  the unnecessary pitch information increases the difficulty of predicting duration. 
 The \emph{Pro-pm} significantly improves pitch accuracy over the \emph{Proposal}, 
 proving the effectiveness of applying pitch metric.
To further explain the reason, we plot the pitch representations of all test utterances output from pitch encoder.
As shown in Figure~\ref{fig:f2}, circle dot denotes pitch representation corresponding to a phoneme
and various colors represent different pitches in order of color depth.
The representations extracted from the \emph{Proposal} and the \emph{Pro-pm} are displayed in the top and the bottom of the figure respectively.
Apparently,
the bottom representations have more compact clusters and fewer outliers than the top.
Only weak superiority of the \emph{Pro-pc} over the \emph{Proposal} may due to  the limited amount of training data and too many pitch categories.
The \emph{Pro-pm-pc} obtains the largest scores in all criteria against other models, demonstrating the more effective representation capability of the two proposed encoders.

\begin{table}[th]
  \caption{Objective evaluation results of different models where Corr. represents the correlation coefficients and Acc. denotes the accuracy.}
  \label{tab:example}
  \centering
  \begin{tabular}{ r@{}c c c c }
    \multicolumn{2}{c}{\textbf{Model}} &  \multicolumn{1}{c}{\textbf{F0 RMSE (Hz)}} &  \multicolumn{1}{c}{\textbf{F0 Corr.}} &  \multicolumn{0}{c}{\textbf{Duration Acc.  (\%)}}\\
   &\emph{Baseline}&34.272&0.833&83.14        \\
   &\emph{Proposal}&32.342&0.853&84.32       \\
   &\emph{Pro-pm}&30.590&0.882&85.47       \\
   &\emph{Pro-pc}&31.691&0.871&85.23        \\
   &\emph{Pro-pm-pc}&\textbf{29.604}&\textbf{0.893}&\textbf{85.71}      \\
  \end{tabular}   
\end{table} 

\subsubsection{Subjective Evaluation}
To compare the subjective performances of the generated singing from different model and ground-truth recording,
the Mean Opinion Score (MOS) is introduced to the auditory test.
Randomly selected 25 music utterances from test set are synthesized by each model.
Then,7 music professionals are asked to grade the MOS score from (bad) to 5 (excellent)
after listening to each audio.
The resulting scores are measured by averaging the MOS of all samples on different comparison models.

\begin{figure}[t]
  \centering
  \includegraphics[width=\linewidth]{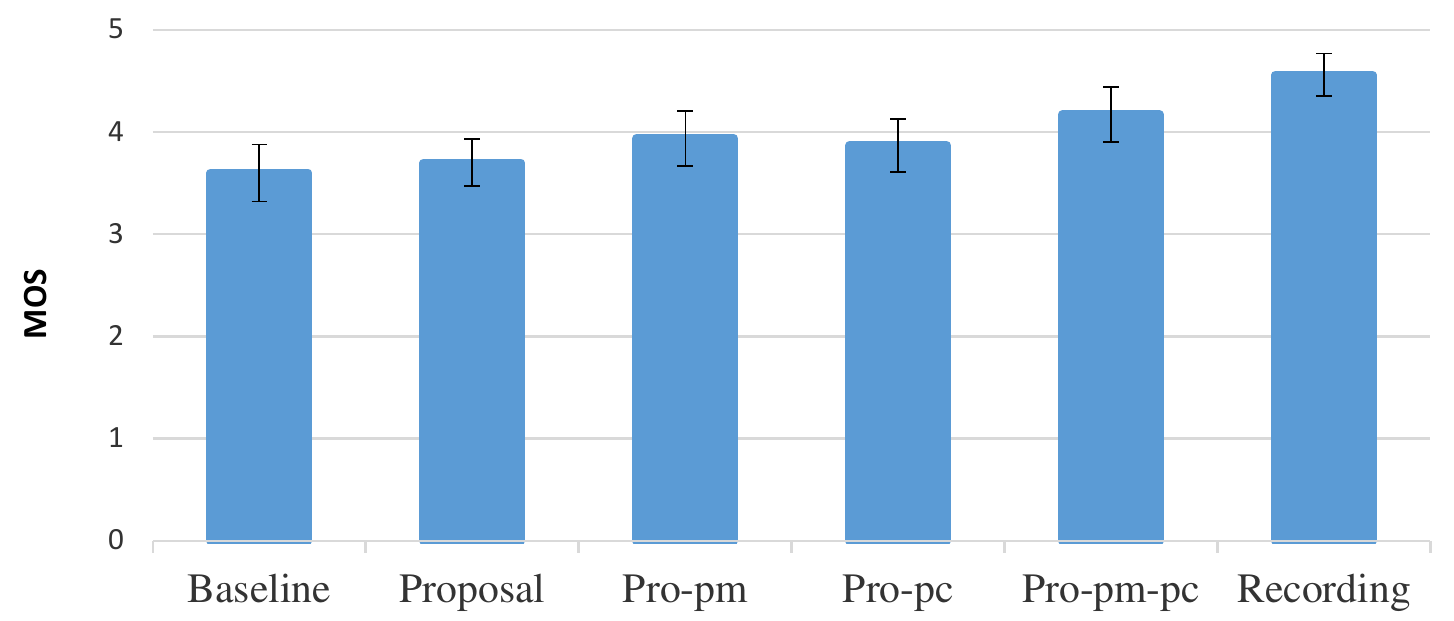}
  \caption{Mean Opinion Score (MOS) test for all systems with 95\% confidence intervals.}
  \label{fig:f3}
\end{figure}

The comparison of the MOS differences between five models and the ground-truth recording is demonstrated in  Figure~\ref{fig:f3}.
The \emph{Baseline} obtains a lowest score, which is far from that of the recording suffering from limited corpus.
The \emph{Proposal} achieve higher score than the \emph{Baseline}, once again proving the rationality of the modified architecture.
The \emph{Pro-pm} and the \emph{Pro-pc} achieve similar scores beyond the \emph{Proposal} achieved because of the more significant encoders adopted.
The biggest gain is achieved by the \emph{Pro-pm-pc} than the other four systems.
In summary, above conclusion are consistent with that of the previous objective evaluation.

\section{Conclusions}

In this paper, a novel  singing voice synthesis system preserving pitch accuracy of generated audio  is proposed.
Two encoders are used to represent phoneme and pitch information respectively.
Further, the phoneme encoder followed by  an adversarially trained pitch classifier is employed to learn pitch-independent phoneme representation.
The pitch encoder is constrained by  a pitch metric loss, which explicitly embeds the distance between adjacent pitches
to enforce pitch representation lying on correct pitch manifolds.
Experimental results demonstrate that the proposed methods obtain higher quality and more accurate pitch of synthesized singing  when limited training corpus is available.
Moreover, the effectiveness of the pitch metric is verified by the analysis that the pitch representations produced by well-trained pitch encoder form good clusters through dimension reduction and visualization.

\bibliographystyle{IEEEtran}
\bibliography{my}
\end{document}